\documentclass[letterpaper,journal]{IEEEtran}
\usepackage{amsmath,amsfonts}
\usepackage{algorithmic}
\usepackage{algorithm}
\usepackage{array}
\usepackage[caption=false,font=normalsize,labelfont=sf,textfont=sf]{subfig}
\usepackage{textcomp}
\usepackage{stfloats}
\usepackage{url}
\usepackage{verbatim}
\usepackage{graphicx}
\usepackage{cite}
\usepackage{amsmath,amssymb,graphicx,bm}
\usepackage{xcolor}
\usepackage{cuted}

\usepackage[normalem]{ulem}

\title{Foil-Mediated Mutual-Impedance in Microwave Cavities with Enhanced Phase Response}

\author{Michael T. Hatzon,~\IEEEmembership{Student Member,~IEEE,} Graeme R. Flower, Robert C. Crew, Ethan Han, Ben T. McAllister, Jeremy F. Bourhill, and Michael E. Tobar,~\IEEEmembership{Fellow,~IEEE}%
\thanks{Michael T. Hatzon, Graeme R. Flower, Robert C. Crew, Jeremy F. Bourhill, and Michael E. Tobar are with the Quantum Technologies and Dark Matter Research Lab, Department of Physics, University of Western Australia, Crawley, WA 6009, Australia (e-mail: michael.hatzon@uwa.edu.au).}
\thanks{Ethan Han, Ben T. McAllister are with the Swinburne University of Technology, Hawthorn, VIC3122, Australia.}}

\begin{document}

\maketitle

\begin{abstract}
We formulate and validate an equivalent circuit model describing mutual coupling between three microwave cavity resonators interconnected via thin metallic foils. Each cavity is represented as a lumped LCR circuit, while the foils act as interfaces that mediate energy exchange via mutual impedance. This coupling mechanism produces interference effects and a controllable antiresonance when the input resonators are amplitude- and phase-balanced, while enabling coupling across a continuous metallic interface without direct antenna or aperture coupling. All three resonators operated in the TM$_{010}$ mode, where two input resonators each excited the third via a thin copper foil. Analytical expressions are derived for the mutual impedance and coupling coefficient of these foils in this geometry. Under balanced conditions, a sharp antiresonance emerges with a several-fold enhanced phase sensitivity at the resonant frequency of the output cavity, consistent with model predictions. The experimentally extracted normalised mutual coupling coefficients, $\Delta_{13}=(10.8\pm0.3)\times10^{-6}$ and $\Delta_{23}=(8.6\pm0.2)\times10^{-6}$, fall within the calculated range $\Delta_{n3}\approx(7\text{--}17)\times10^{-6}$ derived from the foil's electromagnetic properties, where the spread is dominated by the estimated foil thickness uncertainty of $(8.9\pm0.4)\,\mu\mathrm{m}$. These results demonstrate a foil-mediated physical implementation of weak mutual coupling across a metallic interface spanning multiple skin depths, providing a distinct route for engineering controlled interference in three-dimensional multi-resonator systems.
\end{abstract}

\begin{IEEEkeywords}
Microwave resonators, coupled resonators, mutual impedance, dissipative coupling, antiresonance.
\end{IEEEkeywords}

\section{Introduction}
Microwave cavity resonators are core components in microwave engineered systems \cite{Ivanov1998}, oscillators \cite{Locke2008}, and resonator-based sensing and measurement platforms \cite{ebrahimi2019ultrahigh,zhu2025phase}. Their high quality factors and field confinement make them attractive for microwave sensing \cite{ebrahimi2019ultrahigh}, filtering \cite{11433111}, and phase-sensitive measurement \cite{zhu2025phase}. Considerable work has focused on the optimisation, modelling, and engineering of coupling mechanisms in microwave resonators and filters \cite{liu2023exhaustive,yu2022state,zhuo2022parametric}.

Such resonators underpin a wide range of applications, including dielectric and material-loss characterisation \cite{canalias2024transmission,Krupka_1999,Creedon2011,mcrae2020materials}, high-resolution spectroscopy \cite{pozar2021microwave,Farr13,Albertinale2021,Travesedo2025}, cavity electromechanics \cite{Blair95,Cuthbertson96,Aspelmeyer2014}, and precision frequency control \cite{Ivanov2010}.

The way in which two or more resonators are coupled governs the exchange of energy and phase information between them, determining the collective response of the system \cite{hong2004microstrip}. Controlled coupling enables a rich variety of interference effects, including linewidth narrowing, normal-mode splitting, mode hybridisation, or sharp antiresonances \cite{castelli2019normal,teufel2011circuit,ben2026interference,hatzon2024microwave}. By engineering the strength, phase, and balance of inter-resonator coupling, one can tailor the spectral response of the system for specific measurement or filtering purposes, and create conditions for coherent cancellation, abrupt phase variation, and enhanced sensitivity \cite{zhu2022slow}.

Coupled-resonator systems are conventionally realised through shared inductive or capacitive elements, or via aperture-mediated coupling mechanisms \cite{pozar2021microwave}, each producing reactive interactions that modify the system's collective resonance response. Resistive coupling, by contrast, has typically been regarded as a loss mechanism rather than a deliberate means of inter-resonator interaction. Some methods have been developed that invoke resistive coupling, including microwave filter networks \cite{das2018all}, power-line communication systems \cite{swana2015resistive}, and cavity resonances coupling to line resonance due to a finite voltage standing wave ratio (VSWR) \cite{tobar1991generalized}, though these are generally confined to planar or lumped-element implementations. A possible route to implementing such coupling between three-dimensional resonators is through a shared, electrically thin conducting boundary. Thin metallic foils have been employed in other resonant electromagnetic structures, where they influence current distribution and resonant behaviour, and in some cases enable conductive path-mediated interference effects \cite{hu2019experimental, stein2017high}. To our knowledge, thin-foil coupling between 3D microwave cavity resonators, realised here, has not been systematically investigated as a standalone coupling mechanism.

Related dissipative-coupling phenomena have been investigated in microwave cavity-magnonic systems using effective complex coupling descriptions \cite{yang2019control,harder2018level,bourcin2024level}. The characteristic signature of these systems is a level attraction in the strong coupling limit between the two harmonic oscillators, and the phase of the complex coupling rate can be calculated from transverse magnetic field components of the cavity resonance. Here, the coupling between harmonic oscillators is a result of the mutual impedance across the foil, which is derived directly from the foil boundary conditions, and its magnitude is compared quantitatively with the fitted coupling strength.

The motivation for this study arises from a three-cavity microwave system that was engineered with a foil-mediated coupling mechanism for precision measurement \cite{hatzon2025sharp,hatzon2025extreme}. Two input cavities drove a third output cavity through thin copper foils many skin depths thick. When tuned, the system exhibited sharp antiresonances with a strong dependence on the relative input phase.

\begin{figure*}[!t]
\centering
\subfloat[]{
    \includegraphics[width=0.49\textwidth]{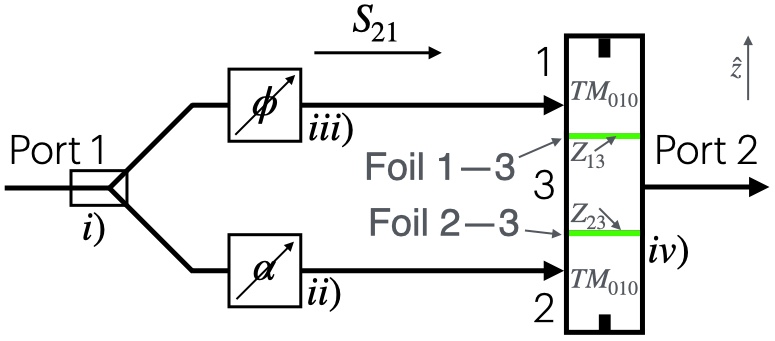}
    \label{simplecircuit}
}
\subfloat[]{
    \includegraphics[width=0.37\textwidth]{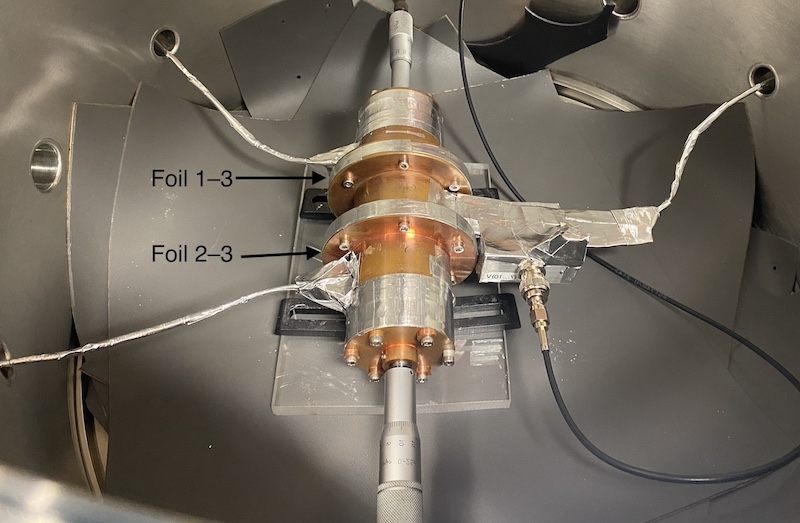}
    \label{photocircuit}
}
\caption{(a) Schematic of the experimental setup, showing 
i) the 3~dB power splitter, 
ii) the variable attenuator, 
iii) the microwave phase shifter, and 
iv) the foil-coupled three-cavity system. 
Resonators 1--3 and 2--3 are coupled through copper foils of thickness $\sim9~\mu\mathrm{m}$, represented by the mutual impedances $Z_{13}$ and $Z_{23}$, respectively. 
(b) Photograph of the fabricated three-cavity apparatus with shielding. The foils are clamped between the resonator volumes at positions labelled `Foil 1--3' and `Foil 2--3'.}
\label{fig:system}
\end{figure*}

Building upon these observations, we enhanced the readout sensitivity and developed an analytical model of the three-cavity system (Fig.~\ref{fig:system}) by introducing mutual impedance terms to describe coupling through the foils. Each cavity is represented as a lumped LCR resonator, with inter-cavity terms proportional to the power transferred across the interface. The model reproduces the observed interference and antiresonant behaviour and is complemented by a first-principles derivation relating the foil's electromagnetic properties to the measured coupling.

While coupling concepts are central to microwave resonator engineering, this work develops and experimentally validates a framework for foil-mediated mutual-impedance coupling between volumetric microwave resonators, with the real part of the mutual impedance describing the dissipative contribution. The resulting framework provides analytical and experimental foundations for modelling this form of inter-resonator coupling and demonstrates its potential for engineering controlled antiresonances and balanced operating conditions in microwave resonator systems.

\section{Experimental Setup and Equivalent Circuit Model}
The experimental schematic is shown in Fig.~\ref{simplecircuit}. A vector network analyser (VNA) measures $S_{21}$ between Port~1 and Port~2. The signal entering Port~1 is split into two paths using a 3~dB power splitter, Fig.~\ref{simplecircuit}, i). One path passes through a variable attenuator ii), acquiring a relative attenuation factor $\alpha$, while the other passes through a microwave phase shifter iii), introducing a relative phase shift $\phi$ between the two input signals. These two signals excite the TM$_{010}$ modes of the external resonators, labelled 1 and 2, in the three-cavity system iv). Both external resonators couple to the central resonator, resonator 3, through copper foils of thickness $(8.9\pm0.4)~\mu\mathrm{m}$. The coupling between resonators 1--3 and 2--3 is represented by the mutual impedances $Z_{13}$ and $Z_{23}$, respectively. Port~2 is connected to a three-stage output amplification chain consisting of two Low Noise Factory amplifiers with approximately 37~dB gain each and one Mini-Circuits amplifier with 17~dB gain, giving a total measured gain of 89.7~dB. After accounting for 4.3~dB of measured component loss (excluding power divider) and an additional estimated cable/connector loss of $1.6 \pm 0.2$ dB, the measured transmission was corrected using a net gain of 83.8~dB.

All measurements presented were acquired with improved electromagnetic shielding and a lower-noise amplification chain, resulting in enhanced contrast and signal-to-noise ratio compared to \cite{hatzon2025sharp,hatzon2025extreme}. Residual fringing fields, leakage paths, and external interference were minimised through the shielding arrangement shown in Fig.~\ref{photocircuit}, including additional sealing of ports and connections, microwave absorber placed around the resonator structure, and enclosure within the conducting shielding chamber. After these measures, the measured response was dominated by the intended TM$_{010}$ modes and the foil-mediated coupling described by the model.

\begin{figure}[!ht]
\centering
	\includegraphics[scale=0.3]{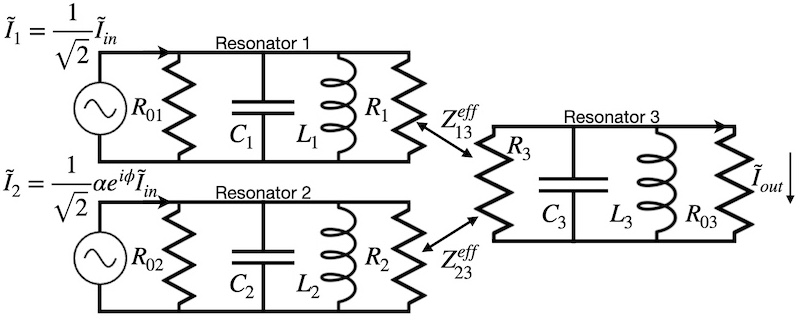}
    \caption{Equivalent-circuit representation of the foil-coupled three-cavity resonator system. Each cavity is represented by a parallel LCR resonator with node voltage $\tilde{V}_n$. The foil-mediated interactions are represented by the effective mutual-coupling terms $Z_{13}^{eff}$ and $Z_{23}^{eff},$ which correspond to the moduli of the underlying complex mutual impedances in the weak-coupling phase convention used for fitting. The input currents $\tilde{I}_1$ and $\tilde{I}_2$ drive resonators 1 and 2, and the output current $\tilde{I}_{out}$ is taken from resonator 3.}
	\label{circuitmodel}
\end{figure}

An equivalent circuit representation of the experimental system is shown in Fig.~\ref{circuitmodel}. Each of the three resonators in Fig.~\ref{simplecircuit} is modelled as an LCR circuit characterised by its resonance frequency $f_{n}$, quality factor $Q_{n}$, and probe coupling coefficients $\beta_{n}$. These modal parameters are related to the circuit elements through the near-resonance impedance of each parallel LCR branch. The conductive foils separating the cavities introduce mutual coupling between the input and output resonators. The LCR representation is used here as a standard single-mode equivalent-circuit model for each cavity near its dominant TM$_{010}$ resonance, and is applied where these modes dominate the measured response. Nearby higher-order modes were sufficiently separated from the measurement band and are not included; the model is therefore restricted to the frequency range over which the three TM$_{010}$ modes dominate.

The coupling between resonators n and 3 is described generally by the complex mutual impedance $Z_{n3}=R_{n3}+iX_{n3},$ where $R_{n3}=Re(Z_{n3})$ is the dissipative component and $X_{n3}= Im(Z_{n3})$ is the reactive component. Writing $Z_{n3}=|Z_{n3}|e^{i \theta_{n3}},$ the common coupling phase may be absorbed into the transmission phase reference, while the relative phase between $Z_{13}$ and $Z_{23}$ may be absorbed into the fitted excitation phase $\phi$ in the present weak-coupling regime. The equivalent circuit used for fitting therefore represents each coupling by the real, non-negative effective value $Z^{eff}_{n3}=|Z_{n3}|.$

In this work, tildes denote peak-amplitude phasors with an $e^{i \omega t}$ time convention. The transfer function, $T$, of the system is defined as
\begin{equation}
	T=\frac{\tilde{I}_{out}}{\tilde{I}_{in}},
    \label{transferfuncdef}
\end{equation} 
where the phasor $\tilde{I}_{in}$ is the Thevenin equivalent input current at port 1, and $\tilde{I}_{out}$ is the current through the port 2 (resonator 3) output resistor. Under matched impedance conditions, this quantity is proportional to the forward transmission coefficient of the microwave circuit, $S_{21}$, and therefore represents the measured complex transmission response. The input current is split into $\tilde{I}_{1}$ and $\tilde{I}_{2}$, and a relative phase, $\phi$, and relative amplitude factor, $\alpha$, is introduced between the two input currents related by \begin{equation}
	\tilde{I}_{1}=\frac{1}{\sqrt{2}}\tilde{I}_{in}
\end{equation} and \begin{equation}
	\tilde{I}_{2}=\frac{1}{\sqrt{2}}\alpha e^{i \phi} \tilde{I}_{in}.
\end{equation}
Each of the resonator equivalent circuits ($n=1,2,3$) has a parallel LC shunt with impedance given by \begin{equation}
    Z_{LC,n}=({i \omega C_n+\frac{1}{i \omega L_n}})^{-1}.
\end{equation} Near resonance, we define $C_n\xrightarrow{}\frac{1}{\omega^2_n L_n }$, $L_{n}\xrightarrow{}\frac{R_n}{\omega_n Q_n}$ and $R_n\xrightarrow{}\beta_n R_{0n}$, for resonator resistance $R_n$, capacitance $C_n$ and inductance $L_n$ at angular frequency $\omega=2 \pi f$. All probes share $R_{01}=R_{02}=R_{03}\equiv R_0$. The first-order resonance expansion becomes \begin{equation}
    Z_{LC,n}\simeq({2\frac{i Q_n}{R_0 \beta_n}(\frac{\omega-\omega_n}{\omega_n})})^{-1},
\end{equation} with $\omega_n = 2 \pi f_n.$ Define the resonator-branch voltages and currents by \begin{equation}
    \mathbf{V=\mathbf{Z_R I_R}},
\end{equation} where \begin{equation}
    \mathbf{V= \begin{pmatrix}
        \tilde{V}_1\\
        \tilde{V}_2\\
        \tilde{V}_3
    \end{pmatrix}}
\end{equation} is the node-voltage vector and \begin{equation}
    \mathbf{I_R}=\begin{pmatrix}
        \tilde{I}_{R1}\\
        \tilde{I}_{R2}\\
        \tilde{I}_{R3}
    \end{pmatrix},
\end{equation} is the current vector associated with the impedance network $\mathbf{Z_R}:$ \begin{equation}
    \mathbf{Z_R}=R_0\begin{pmatrix}
        \beta_1 & 0 & \sqrt{\beta_1 \beta_3} \Delta_{13}\\
        0 & \beta_2 & \sqrt{\beta_2 \beta_3} \Delta_{23}\\
        \sqrt{\beta_1 \beta_3} \Delta_{13} & \sqrt{\beta_2 \beta_3} \Delta_{23} & \beta_3
    \end{pmatrix}.
\end{equation}  The dimensionless effective coupling strength used in this phase-normalised representation is defined as \begin{equation}
    \Delta_{n3}=\frac{Z_{n3}^{eff}}{\sqrt{R_n R_3}}=\frac{|Z_{n3}|}{\sqrt{R_n R_3}}=\frac{|Z_{n3}|}{R_0 \sqrt{\beta_n \beta_3}}.
    \label{mutual_coupling}
\end{equation} Thus, $\Delta_{n3}$ describes the magnitude of the effective coupling and does not assume that $Z_{n3}$ is purely resistive. Corrections that depend explicitly on the coupling phase enter the transfer function denominator at order $\Delta_{n3}^2$ and are negligible for the low coupling strengths considered here. The three nodal equations become the vector equation \begin{equation}
    \mathbf{I_s}=\begin{bmatrix}
        \text{diag}(\frac{1}{R_0}+\frac{1}{Z_{LC,1}},\frac{1}{R_0}+\frac{1}{Z_{LC,2}},\frac{1}{R_0}+\frac{1}{Z_{LC,3}})+\mathbf{Z_R^{-1}}\end{bmatrix}\mathbf{V},
\end{equation} with the current source term \begin{equation}
    \mathbf{I_s}=\frac{\tilde{I}_{in}}{\sqrt{2}}\begin{pmatrix}
        1\\
        \alpha e^{i \phi}\\
        0
    \end{pmatrix}.
\end{equation} After solving this vector equation for the voltages, one can construct the output current $\tilde{I}_{out}=\frac{\tilde{V}_3}{R_{03}}$. Recalling Eq.~\eqref{transferfuncdef}, the transfer function can be written as \begin{strip}
\hrule
    \begin{equation}
    T=\frac{\sqrt{\beta_3}(\sqrt{\beta_1}\gamma_2 \Delta_{13}+\alpha e^{i \phi}\sqrt{\beta_2} \gamma_1 \Delta_{23})}{\sqrt{2}(\gamma_1 \gamma_2 \gamma_3 - \gamma_2(\gamma_1-1)(\gamma_3-1)\Delta_{13}^2-\gamma_1(\gamma_2-1)(\gamma_3-1)\Delta_{23}^2)},
\label{fulltransferfun}
\end{equation}
\hrule
\end{strip} where we have introduced the dimensionless resonant factor \begin{equation}
    \gamma_n(f)=1+\beta_n+2iQ_n\frac{f-f_n}{f_n}.
\end{equation} In the small mutual-coupling limit where $\Delta_{n3}^2\xrightarrow{}0$, such as the regime of the present study, Eq.~\eqref{fulltransferfun} becomes \begin{equation}
    T\approx\frac{\sqrt{\beta_3}(\sqrt{\beta_1}\gamma_2 \Delta_{13}+\alpha e^{i \phi}\sqrt{\beta_2} \gamma_1 \Delta_{23})}{\sqrt{2}\gamma_1 \gamma_2 \gamma_3}.
\end{equation}

Interesting phenomena arise when the contributions to the output current are closely balanced. To determine the balance condition, we set the attenuation factor $\alpha$ acting on $\tilde{I}_{2}$, such that $T=0$. Under the assumption that all three resonators are tuned to the same resonant frequency ($f_{1}=f_{2}=f_{3}$), and that the relative phase is matched, the required attenuation is, 
\begin{equation}
	\alpha=\frac{\sqrt{\beta_{1}}(1+\beta_{2})\Delta_{13}}{\sqrt{\beta_{2}}(1+\beta_{1})\Delta_{23}}.
    \label{balance}
\end{equation}
Eq.~\eqref{balance} provides the symmetric-limit intuition; in practice small detunings shift the optimal balance point slightly, which is captured in the full model (Eq.~\eqref{fulltransferfun}) used for fitting.

\section{Fitting the Data}
\begin{figure}[h!]
    \centering
    \subfloat[]{%
        \includegraphics[width=\linewidth]{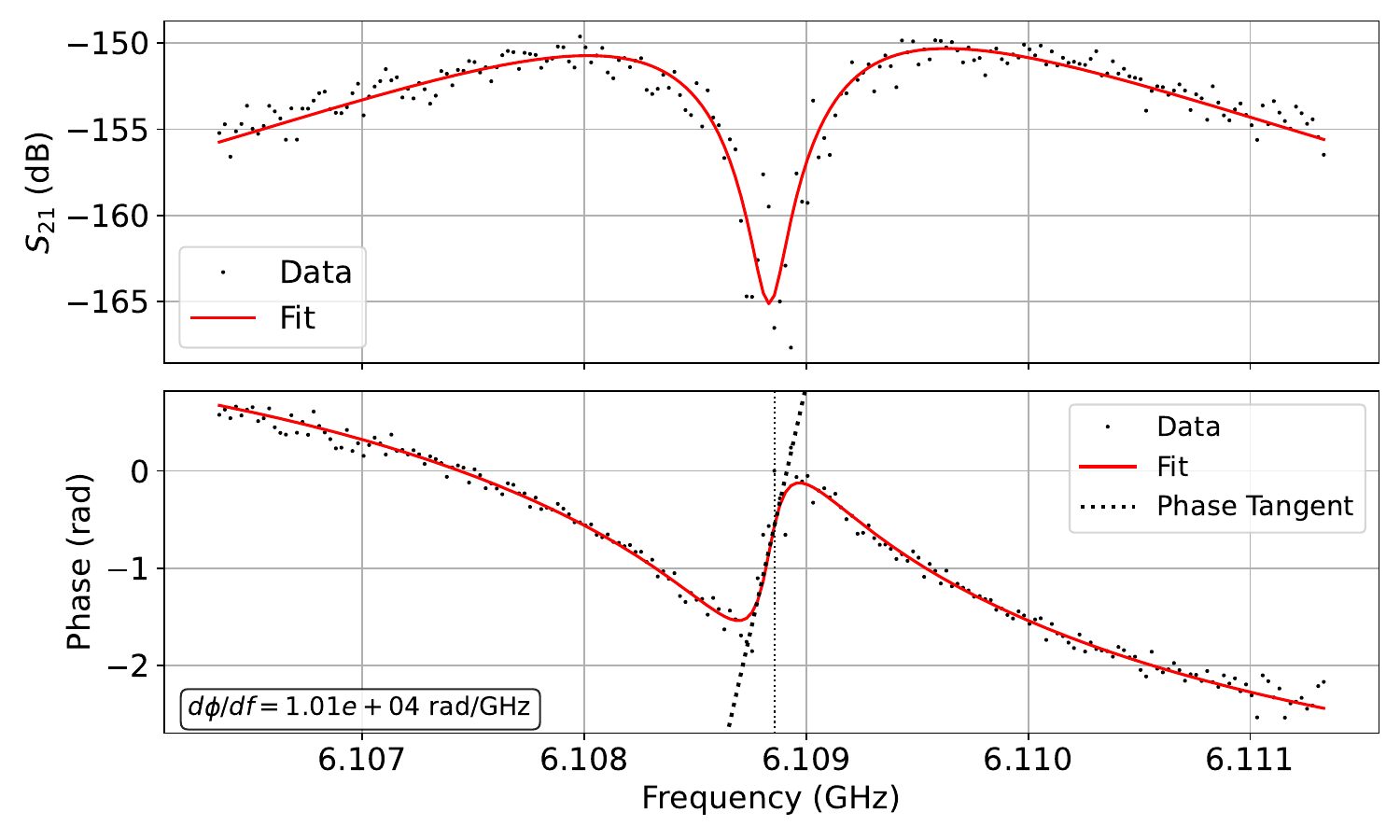}%
        \label{fig:Fig1a}}
    \vfill
    \subfloat[]{%
        \includegraphics[width=\linewidth]{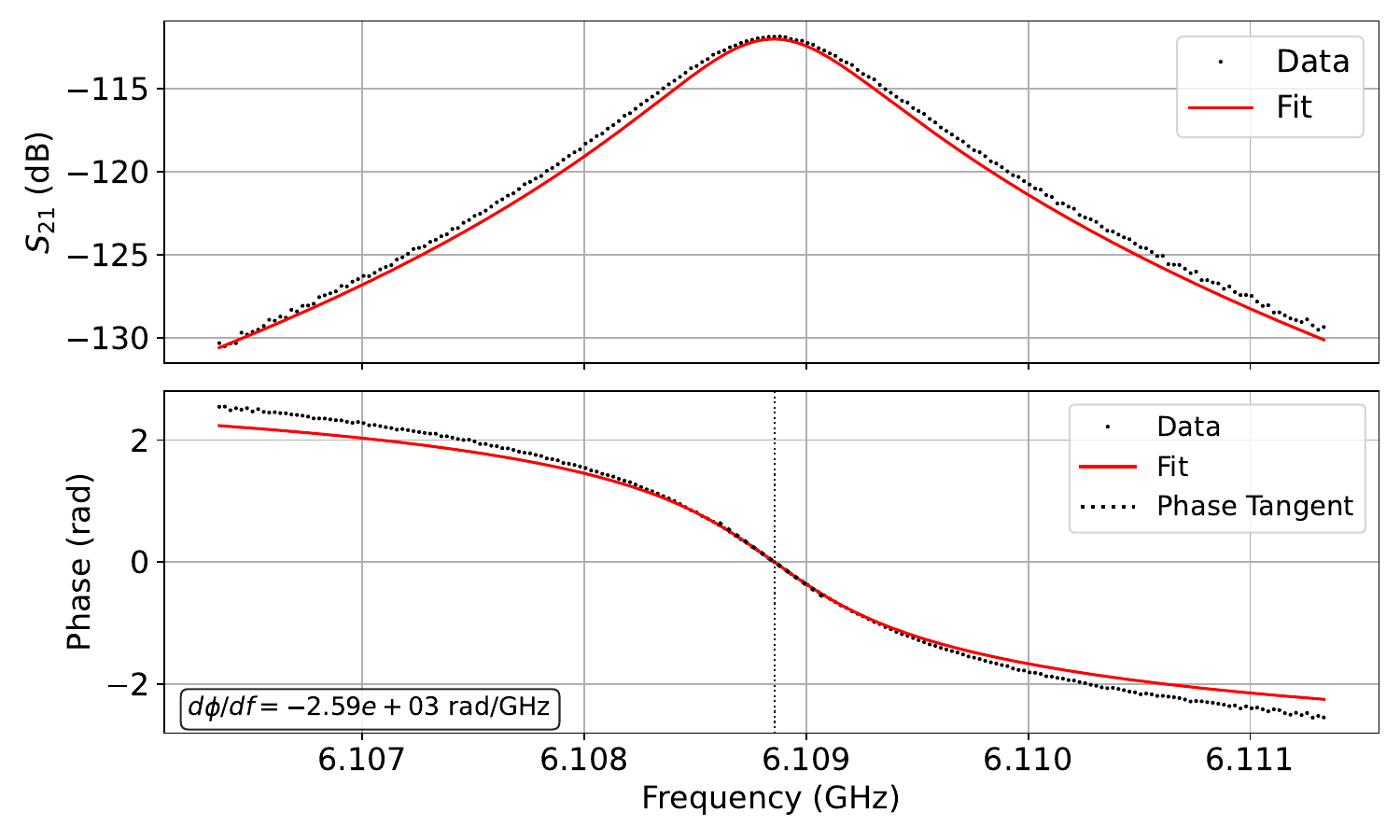}%
        \label{fig:Fig1b}}
    \vfill
    \subfloat[]{%
        \includegraphics[width=\linewidth]{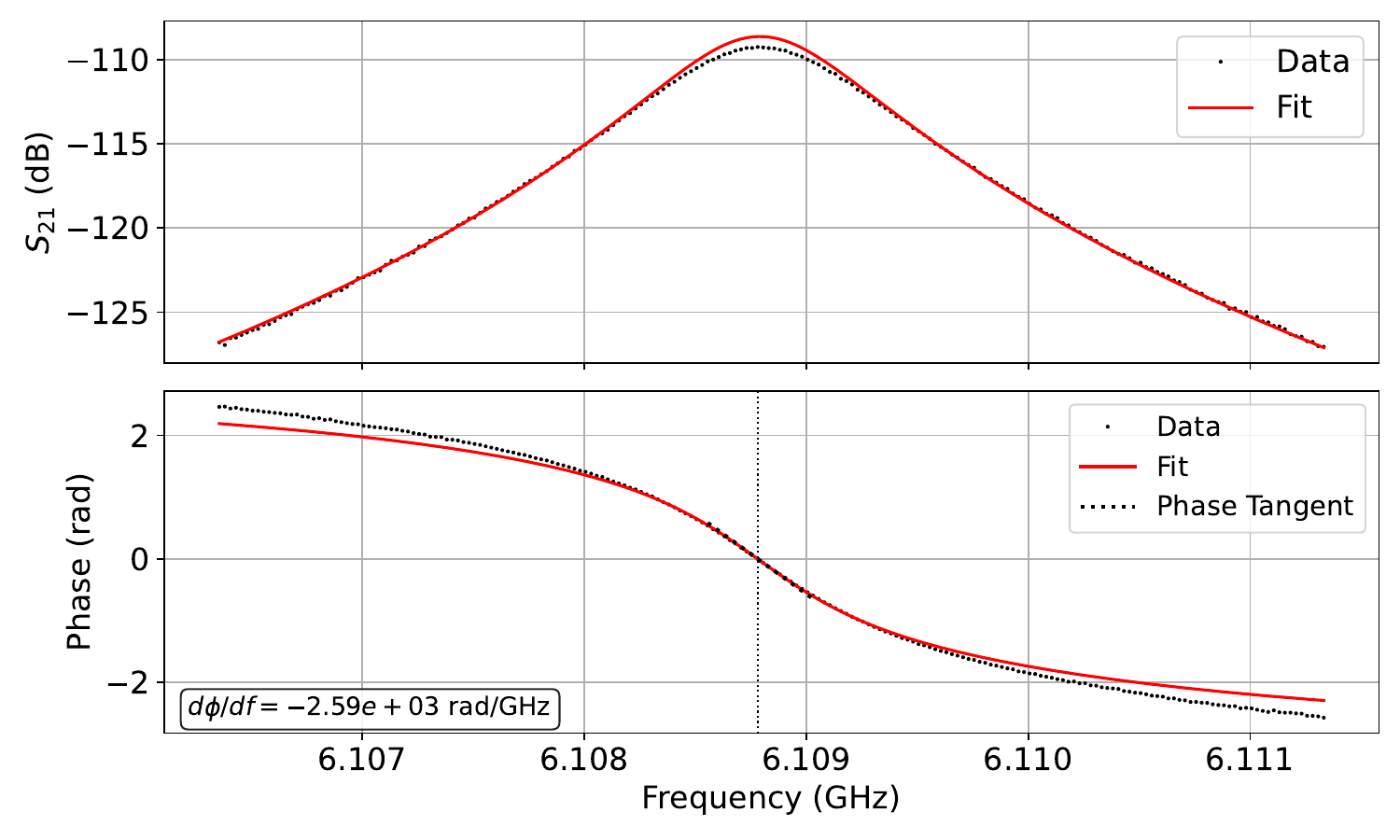}%
        \label{manysharp}}
    \caption{Measured magnitude and phase responses with the best-fit transfer function from Eq.~\eqref{fulltransferfun}. The corresponding fit parameters are listed in Table~\ref{fitparams}. Panels (a)--(c) correspond to phase-shifter settings of 4.90, 2.45, and 0.00 mm, respectively, with the relative excitation phases constrained by the measured phase-shifter calibration. These settings span approximately 2.96 rad, from near $2\pi$ to near $\pi$. The input power at Port~1 was 20~dBm, and the transmission data were corrected using a net measurement-chain gain of 83.8 dB.}
    \label{fits}
\end{figure}

Three experimental spectra are plotted in Fig.~\ref{fits} and fitted using the transfer function in Eq.~\eqref{fulltransferfun} via standard nonlinear least-squares optimisation in Python. The resonance frequencies $f_1$ and $f_2$, probe coupling coefficients $\beta_1$, $\beta_2$ and $\beta_3$, relative attenuation $\alpha$, and overall amplitude scale were fixed using independent measurements or calibration. The quality factors $Q_1$, $Q_2$ and $Q_3$, together with the coupling coefficients $\Delta_{13}$ and $\Delta_{23}$, were shared across all three spectra. A separate $f_3$ and constant transmission-phase offset were fitted for each spectrum. A common phase zero-point $\phi_0$ was fitted, while the relative phase progression was constrained by the phase-shifter calibration, $\phi_i=\phi_0-(\frac{171.6^\circ }{4.95 mm})x_i,$ where $x_i=0,2.45,4.90$ mm. This gives the relative phase offsets of $0^\circ,-84.93^\circ,-169.87^\circ,$ respectively.

The parameter extraction was constrained by independently measured quantities wherever possible, rather than performed as an unconstrained fit of all model parameters. The parameters were obtained by jointly fitting the real and imaginary components of the complex transmission response. A frequency-independent phase-reference offset was fitted separately for each spectrum to account for arbitrary changes in the measured transmission-phase reference. These offsets do not alter the calibrated relative excitation-phase progression. Each spectrum's residual was normalised by its maximum measured transmission magnitude so that the three datasets contributed comparably to the global optimisation. Robustness checks over a broad range of initial parameter values and fitting bounds showed that good fits reproducing the measured interference behaviour did not occur generically; parameter sets far from the reported weak-coupling regime failed to reproduce the observed cancellation and phase evolution. The extracted values should therefore be interpreted as effective single-mode parameters over the measured frequency range.

There is good agreement between the model and experiment, with the extracted parameters summarised in Table~\ref{fitparams}. The model reproduces this measured dependence on the relative input phase $\phi$: as $\phi$ is tuned towards the balanced condition, a sharp antiresonance forms and the phase slope steepens.

Small system asymmetries (e.g., unequal resonance frequencies or $Q$-factors) create the ideal single antiresonance when $\phi$ approaches $\pi$, as observed in Fig.~\ref{fig:Fig1a}. Additional imbalance in $\alpha$ can lead to a splitting of this feature within the nearby phase region. In the symmetric limit, where all relevant parameters are equal, exact cancellation occurs at $(\phi=\pi)$ and broadband suppression is recovered across the plotted frequency span.

\begin{figure*}[t]
    \centering
    \begin{minipage}{0.48\textwidth}
        \centering
        \includegraphics[width=\linewidth]{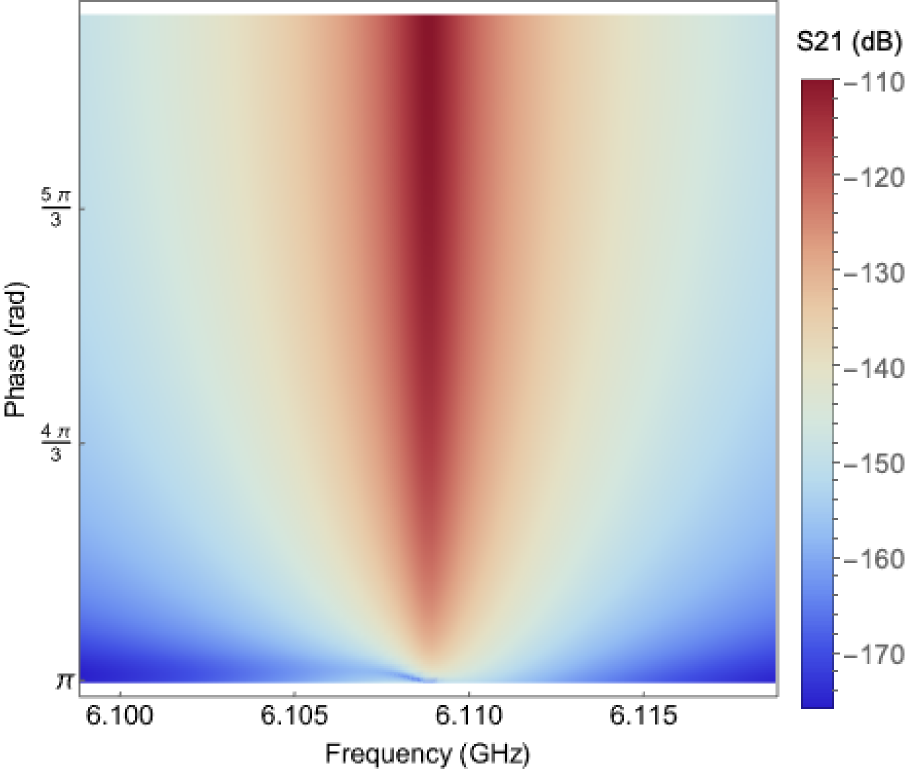}
        \par\smallskip{\normalsize (a) Model\par}
    \end{minipage}\hfill
    \begin{minipage}{0.48\textwidth}
        \centering
        \includegraphics[width=\linewidth]{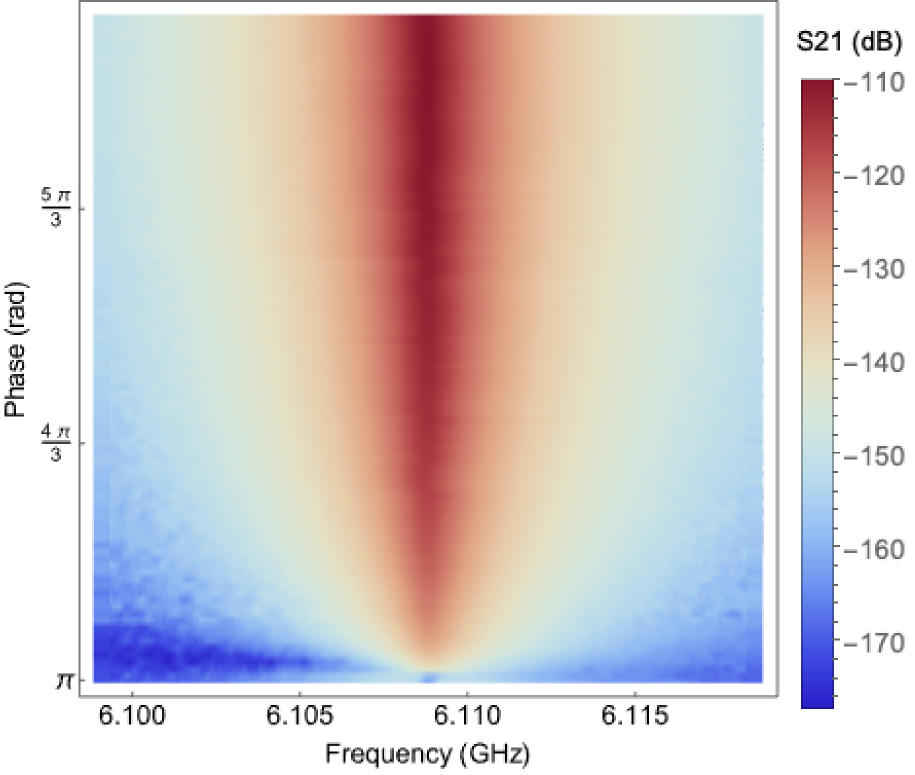}
        \par\smallskip{\normalsize (b) Data\par}
    \end{minipage}  
    \caption{Comparison between the (a): simulated transfer function (using the parameters in Table~\ref{fitparams}) for a near-$\pi$ span of $\phi$, and (b): the Savitzky-Golay smoothed (m=20, order=4) 98 measured spectra. The absolute phase reference is arbitrary; therefore a constant phase offset was applied for visual alignment between model and data. Both plots are shown for the experimentally determined $\phi$ range.}

    \label{fig:model_vs_data}
\end{figure*}

The central resonator frequency, $f_3$, was allowed to vary within a narrow range during the fitting procedure, substantially smaller than the resonator bandwidth (of order MHz). This small freedom accounts for slight frequency drift observed during measurements, most likely arising from temperature or pressure fluctuations that become visible at the achieved low-noise measurement level. $f_3$ and a constant phase-reference offset were allowed to vary independently between spectra. The relative phase excitations were linked by the measured phase shifter calibration through a single fitted phase zero-point. All remaining fitted parameters were shared between the three spectra. The model shows strong agreement with measurements across the full tuning range, as demonstrated in Fig.~\ref{fig:model_vs_data}, which compares the measured response with the simulated transfer function as $\phi$ is swept over approximately 2.96 rad, consistent with the calibrated phase-shifter progression at 6.1~GHz.

\begin{table}[h!]
\centering
\caption{System parameters used for the fits in Fig.~\ref{fits}.}
\renewcommand{\arraystretch}{1.25}
\begin{tabular}{|l|c|}
\hline
\multicolumn{2}{|c|}{\textbf{Measured parameters}} \\
\hline\hline
$f_1$ (GHz) & $6.10883$ \\
$f_2$ (GHz) & $6.10886$ \\
\hline
$\beta_1$ & $0.81$ \\
$\beta_2$ & $1.26$ \\
$\beta_3$ & $0.66$ \\
\hline
$\alpha$ & $1.26$ \\
\hline
\multicolumn{2}{|c|}{\textbf{Fitted parameters}} \\
\hline\hline
$f_{3},0.00$ mm (GHz) & $6.10879$ \\
$f_3,2.45$ mm (GHz) & $6.10885$ \\
$f_3, 4.90$ mm (GHz) & $6.10886$ \\
\hline
$Q_1$ & $2.7\times10^3$ \\
$Q_2$ & $3.1\times10^3$ \\
$Q_3$ & $10.8\times10^3$ \\
\hline
$\Delta_{13}$ & $(10.8\pm0.3)\times10^{-6}$ \\
$\Delta_{23}$ & $(8.6\pm0.2)\times10^{-6}$ \\
\hline
$\phi_0$ (rad) & $-0.188$\\
\hline
\end{tabular}
\label{fitparams}
\end{table}

The two input resonators exhibit comparable quality factors, whilst the output resonator has a higher quality factor, and are only weakly detuned, consistent with the near-symmetric behaviour expected for the balanced interference condition. The fitted mutual coupling coefficients are very small, on the order of $10^{-6}$, consistent with the weak interaction expected when the electromagnetic field couples across many skin depths of copper. The small difference between $\Delta_{13}$ and $\Delta_{23}$ likely arises from variations in surface field distributions, minor geometric asymmetries, or small differences in foil thickness. In Section~\ref{Coupling}, the physical interpretation of this mutual coupling is discussed and a predicted value for the coupling strength based on the foil-mediated mutual impedance is derived.

Uncertainties in the fitted coupling coefficients were evaluated using a Monte Carlo refitting procedure with $10^4$ trials. In each trial, a common loss perturbation ($\sqrt{2} \times 0.14=0.20$ dB), obtained by combining two independent 0.14 dB cable-loss uncertainties in quadrature, was applied to all three datasets. The independently measured $f_1$ and $f_2$ were also sampled from a normal distribution with standard deviation 56 kHz. The complete global least-squares fit was then repeated with these samples. Only the independently characterised measurement-chain loss and $f_1$, $f_2$ uncertainties were sampled, as measurement precision identified these as the dominant external uncertainty sources. All fitted parameters were re-optimised every trial, while other inputs were held at their nominal values. The uncertainties of $\Delta_{13}$ and $\Delta_{23}$ are derived from the standard deviations of the refitted distributions, giving $\Delta_{13}=(10.8\pm0.3)\times 10^{-6}$ and $\Delta_{23}=(8.6\pm0.2)\times 10^{-6},$ and represent the propagation of the dominant calibration uncertainties.

\begin{figure}[t]
    \centering
    \includegraphics[width=0.99\columnwidth]{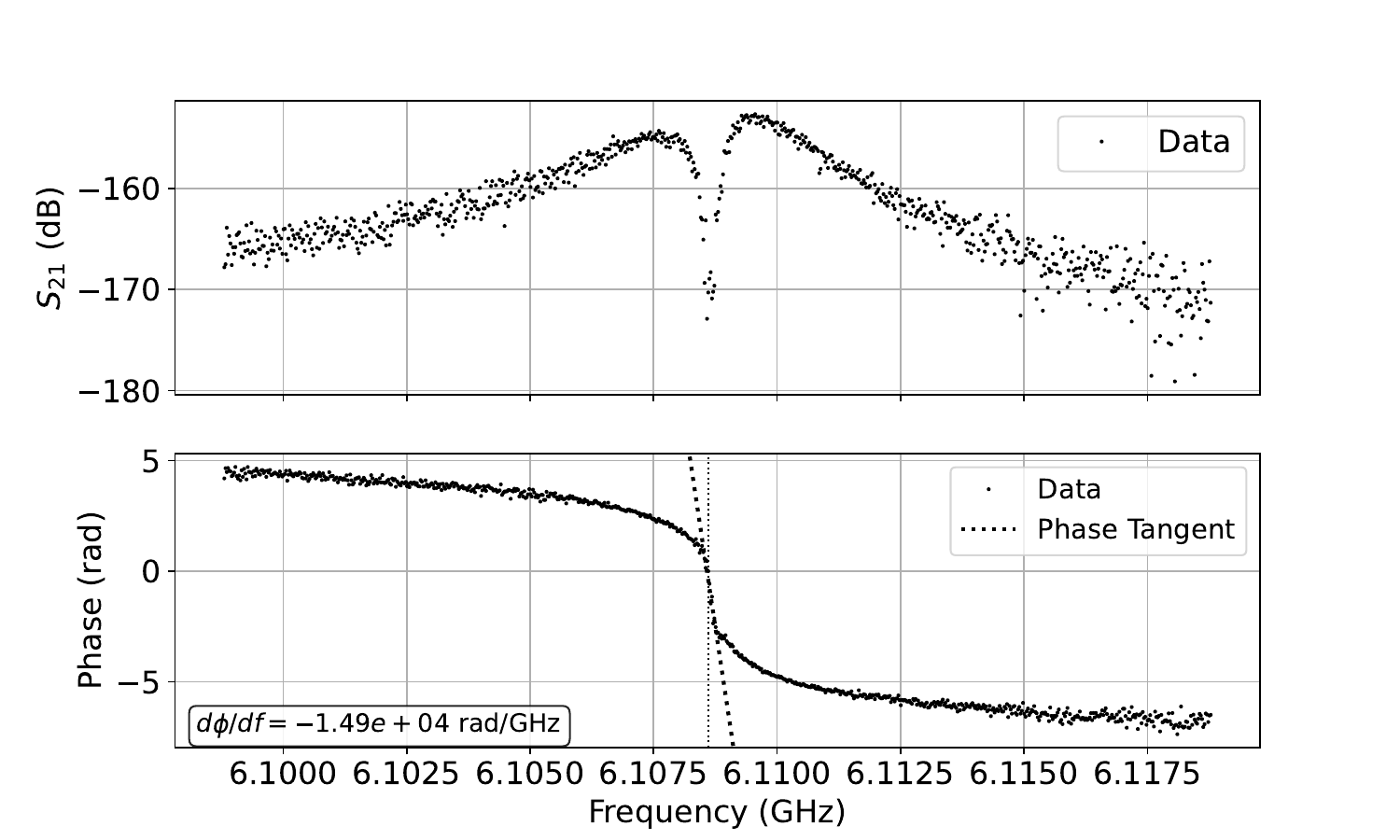}
    \caption{Measured magnitude and phase response near the balanced interference condition showing enhanced phase sensitivity. The local phase slope is enhanced several-fold relative to the near-resonant operating point shown in Fig.~\ref{manysharp}, consistent with the behaviour expected near the antiresonance condition.}
    \label{fig:phase_slope_enhancement}
\end{figure}

To directly measure the phase enhancement, an independent spectrum in Fig.~\ref{fig:phase_slope_enhancement} was acquired by finely tuning the balance parameters ($\phi$ and $\alpha$) to approach the destructive condition. This demonstrates a several-fold increase in the local phase slope compared with the unbalanced configuration. This provides a direct, data-driven confirmation of the phase enhancement predicted by the model. Though both Fig.~\ref{fig:phase_slope_enhancement} and Fig.~\ref{fig:Fig1a} show the destructive case, small differences in balance parameters can cause the polar plot to circle the origin and give the enhanced phase response in Fig.~\ref{fig:phase_slope_enhancement}, rather than the inversion seen in Fig.~\ref{fig:Fig1a}. This is similar to the difference between an under-coupled and over-coupled resonator.

Due to ambient temperature and pressure fluctuations, maintaining the antiresonant condition is experimentally challenging, as the balanced interference point is extremely sensitive to small frequency drifts. Improved thermal and environmental stability, such as operation under vacuum or at cryogenic temperature, is expected to significantly improve the robustness of this operating regime and may be explored in future work.

Having established the phenomenology of the enhancement, we next consider the physical origin of the weak mutual coupling responsible for this behaviour.

\section{Coupling Across a Metallic Foil}
\label{Coupling}
\subsection{Derivation}
\label{couplingderivation}

To connect the fitted coupling strength $\Delta_{n3}$ with the electromagnetic properties of the foil, we derive the complex mutual impedance $Z_{n3}$ from first principles for the equivalent circuit shown in Fig.~\ref{circuitmodel}. Its modulus $|Z_{n3}|$ determines the effective coupling strength defined in Eq.~\eqref{mutual_coupling}, while its real part $R_{n3}=Re(Z_{n3})$ describes the dissipative component of the interaction.

First, consider the electric field induced at either side of a thin sheet, $\tilde{E}_{n}$ and $\tilde{E}_{3}$, for foil sides n (where n is now either 1 or 2, representing the side of the associated input resonator) and 3, respectively: \begin{eqnarray}
    \label{En}
	\tilde{E}_{n}=Z_{s} \tilde{J}_{sn} + Z_{t} \tilde{J}_{s3}, \quad \\
    \label{E3}
	\tilde{E}_{3}=Z_{s} \tilde{J}_{s3} + Z_{t} \tilde{J}_{sn}, \quad
\end{eqnarray} with surface currents $\tilde{J}_{sn}$ and $\tilde{J}_{s3}$ at sides n and 3, respectively, self impedance $Z_{s}$, at the two sides, and $Z_{t}$ is the transfer impedance. The expression for $Z_{t}$ is obtained from a transition boundary condition\cite{COMSOL_Optics_UserGuide_5.5}, which models a geometrically thin, but not electrically thin, sheet. To our knowledge, such a derivation is not readily available in the literature, and is thus provided in Appendix~\ref{derivation}. Here, the self and transfer impedances are defined as \begin{eqnarray}
	Z_{s}=-\frac{i \omega \mu}{k} \frac{1}{tan(kd)}, \quad \\
	Z_{t}=-\frac{i \omega \mu}{k} \frac{1}{sin(kd)}, \quad
\end{eqnarray} with propagation distance (or foil thickness) d. The plane wave propagation constant $k$, is given by $k=\omega \sqrt{(\epsilon+(\frac{\sigma}{i \omega}))\mu}$, with angular frequency $\omega$, conductivity $\sigma$, and the media's permittivity and permeability $\epsilon$ and $\mu$, respectively. We retain the complete complex mutual impedance throughout the derivation. Its modulus is used for comparison with the fitted effective coupling strength.

The total complex power delivered to the two foil surfaces is \begin{equation}
\label{pdiss}
P=\frac{1}{2}\int_{S_f} \tilde{E}_n.\tilde{J}_{sn}^{*}+\tilde{E}_{3}.\tilde{J}_{s3}^*dS,	\quad
\end{equation} where $S_f$ denotes the common geometric area of the two foil faces. After substituting Eqs.~\eqref{En} and \eqref{E3}, this becomes \begin{equation}
P=\frac{1}{2} \int_{S_f} Z_{s} |\tilde{J}_{sn}|^{2}+Z_{s}|\tilde{J}_{s3}|^{2}+2Z_{t}Re(\tilde{J}_{sn}.\tilde{J}_{s3}^{*})dS, \quad
\end{equation} where the final term is the mutual power \begin{equation}
\label{diss1}
P_{n3}= Z_{t}\int_{S_f} Re(\tilde{J}_{sn} . \tilde{J}_{s3}^{*})dS. \quad
\end{equation}
This can also be written as \begin{equation}
	\label{diss2}
	P_{n3}=Z_{n3} Re(\tilde{I}_{Rn} \tilde{I}_{R3} ^*)
\end{equation} with the equivalent resonator-branch current phasors $\tilde{I}_{Rn}$ and $\tilde{I}_{R3}$ at sides n and 3, respectively, and in terms of the mutual impedance $Z_{n3}$. Combining these two mutual-power expressions gives \begin{equation}
Z_{n3}= Z_{t}\frac{\int_{S_f} Re(\tilde{J}_{sn} . \tilde{J}_{s3}^{*})dS}{Re(\tilde{I}_{Rn} \tilde{I}_{R3}^{*})}.
\label{pretransadmittance}
\end{equation} To evaluate this expression, separate the common phase associated with each resonator: $\tilde{J}_{sj}=\vec{J}_{sj}e^{i \phi_j}$, $\tilde{I}_{Rj}=\frac{\tilde{V}_{j}}{Z_j},$ $\tilde{V}_j=\sqrt{\frac{2 U_j}{C_j}}e^{i \phi_j},$ where $j=n,3$. Inserting these, Eq.~\eqref{pretransadmittance} becomes 
\begin{equation}
\label{transadmittance}
Z_{n3}=Z_{t} \frac{\int_{S_f} Re(\vec{J}_{sn} . \vec{J}_{s3}^{*}e^{i(\phi_n-\phi_3)})dS}{2Re( \frac{1}{Z_n Z_3^*}\sqrt{\frac{U_{n} U_{3}}{C_{n} C_{3}}}e^{i(\phi_{n}-\phi_{3})})}. \quad
\end{equation} Here, $Z_j$, $C_j$, $U_j$ and $\phi_j$ denote the resonator impedance, modal capacitance, stored energy, and voltage phase, respectively, for $j \in \{n,3\}$. $\tilde{J}_{sn}$ and $\tilde{J}_{s3}$ can be related to the magnetic fields at their respective surfaces, $\tilde{H}_{tj}=\vec{H}_{tj}e^{i\phi_j}$, giving \begin{equation}
\label{admittance}
Z_{n3}=Z_{t} \frac{\int_{S_f} Re((\vec{n}_{n}\times \vec{H}_{tn}).(\vec{n}_{3} \times \vec{H}_{t3})^{*} e^{i(\phi_n-\phi_3)})dS}{2Re(\frac{1}{Z_n Z_3^*}\sqrt{\frac{U_{n} U_{3}}{C_{n} C_{3}}}e^{i(\phi_{n}-\phi_{3})})},	\quad
\end{equation} with $\vec{n}_{n}$ and $\vec{n}_{3}$ unit vectors normal to their respective sides in the $\pm z$ directions. By noting that the dot product of the two cross terms is purely real for the TM$_{010}$ mode, and by observing that we are interested in the behaviour near resonance ($Z_{n},Z_3\xrightarrow{}R_n,R_3)$, the expression simplifies to \begin{equation}
Z_{n3}=\frac{1}{2}\sqrt{\frac{C_n C_3}{U_n U_3}}Z_{t} R_n R_3 \int_{S_f} Re((\vec{n}_{n}\times \vec{H}_{tn}).(\vec{n}_{3} \times \vec{H}_{t3})^{*})dS.
\label{unsubbedimpednace}
\end{equation} $R_{n}$ and $R_{3}$ are the parallel shunt resistance of the respective resonator given by \begin{equation}
\label{cavityparallelresistance}
R_{j}=\frac{h_{j}^{2} \mu_{0}}{2 \pi a_{j} (a_{j}+h_{j}) R_{surface} \epsilon_{0} J_{1}(\chi_{01})^{2}}.
\end{equation} Here, $j\in \{n,3\},$ $a_j$ and $h_j$ are the radius and height of cavity $j$, and $R_j$ is its parallel shunt resistance. $\mu_{0}$ is the permeability of free space, and $R_{surface}$ is the surface resistance $R_{surface}=\sqrt{\frac{\omega \mu_{0} \rho}{2}}$ with resistivity $\rho$. Near both sides of the foil, the tangential magnetic field of a cylindrical resonator operating in the TM$_{010}$ is 
\begin{equation}
    \label{mag}
\vec{H}_{t}=i E_0 \sqrt{\frac{\epsilon_0}{\mu_0}} \begin{pmatrix}
	0 \\ J_{1}(\frac{\chi_{01} r}{a}) \\ 0
\end{pmatrix},
	\quad
\end{equation}
at a radius $r$, with real electric field normalisation constant $E_{0}$ and cavity radius $a$ for the respective resonator. $\vec{H}_{tn}$ and $\vec{H}_{t3}$ involve the microwave experimental parameters for their respective sides. $J_{m}$ is the Bessel function of the first kind of order m, $J_{m}'$ denotes its derivative, and $\chi_{mu}$ is the u-th positive zero. This mode has capacitance \begin{equation}
\label{capacitance}
C=\frac{\pi \epsilon_{0} a^{2} J_{1}(\chi_{01})^{2}}{h} \quad
\end{equation}
and stored energy \begin{equation}
	\label{stored}
U=\frac{1}{2}\pi \epsilon_{0} a^{2} h E_{0}^{2} J_{1}(\chi_{01})^{2},	\quad
\end{equation} with cavity height $h$ and permittivity of free space $\epsilon_{0}$. Since, in this case, the metallic foil couples the ends of two cylindrical resonators of different radii, the integral is computed over the smaller radius ($a_n < a_3$). Substituting Eqs.~\eqref{mag}--\eqref{stored} into \eqref{admittance} the mutual impedance becomes \begin{equation}
\label{mutualadmittance}
Z_{n3}=2 \pi \epsilon_0 \frac{Z_{t} R_n R_3 a_{n}^{3} a_{3} J_{0}(\frac{a_{n} \chi_{01}}{a_{3}})J_{1}(\chi_{01})}{\chi_{01} \mu_0 h_{n} h_{3} (a_{n}^{2}-a_{3}^{2})}.
\end{equation} The subscript on $a$ and $h$ identify the radii and height of the corresponding cavity. The dissipative component of this mutual impedance is \begin{equation}
\label{mutualresistanceeq}
R_{n3}=2 \pi \epsilon_0 \frac{Re(Z_{t}) R_n R_3 a_{n}^{3} a_{3} J_{0}(\frac{a_{n} \chi_{01}}{a_{3}})J_{1}(\chi_{01})}{\chi_{01} \mu_0 h_{n} h_{3} (a_{n}^{2}-a_{3}^{2})}.
\end{equation}

Eq.~\eqref{mutualresistanceeq} is obtained using the lumped representation of Fig.~\ref{circuitmodel}, in which the currents associated with the foil-mediated interaction are represented by the equivalent resonator currents. More precisely, the transition boundary condition is defined in terms of the distinct currents flowing on the two faces of the foil. Appendix \ref{Appendix:correction} introduces these currents in a more rigorous separate-branch formulation and shows that, in the weak-coupling and near-resonance limits, it reduces to the lumped model.

\subsection{Calculated Coupling Strength}
The physical parameters of the experiment are listed in Table \ref{physicalconstants} and are substituted into Eq.~\eqref{mutualadmittance} to calculate the complex mutual impedance. The theoretical quantity corresponding to the fitted parameter in Eq.~\eqref{mutual_coupling} is the normalised mutual-impedance modulus \begin{equation}
    \Delta_{n3}^{th}=\frac{|Z_{n3}|}{\sqrt{R_n R_3}}.
    \label{theory}
\end{equation}

\begin{table}[!ht]
	\centering
	\caption{Physical constants and parameters of the experiment.}
	\label{physicalconstants}
	\begin{tabular}{ll}
		\hline
		\textbf{Parameter} & \textbf{Value} \\
		\hline
        $f$     & $\approx 6.109$ \text{GHz} \\
		$a_{n}$ & $(18.00\pm0.05) \times 10^{-3}\ \text{m}$ \\
		$a_{3}$ & $(18.73\pm0.05) \times 10^{-3}\ \text{m}$ \\
		$h_{n}$ & $(35.00\pm0.025) \times 10^{-3}\ \text{m}$ \\
		$h_{3}$ & $(38.64\pm0.025) \times 10^{-3}\ \text{m}$ \\
        $d$ & $(8.9\pm 0.4)\times 10^{-6}\ \text{m}$ \\
		$\varepsilon_{0}$ & $8.854 \times 10^{-12}\ \text{F/m}$ \\
		$\mu_{0}$ & $4\pi \times 10^{-7}\ \text{H/m}$ \\
		$\rho_{\text{Cu}}$ & $1.68 \times 10^{-8}\ \Omega\,\text{m}$ \\
		$\sigma_{\text{Cu}}$ & $5.96 \times 10^{7}\ \text{S/m}$ \\
		\hline
	\end{tabular}
\end{table}

The uncertainty in the theoretically predicted coupling was propagated using a Monte Carlo simulation. Because the predicted coupling depends strongly and nonlinearly on foil thickness, its dominant uncertainty was propagated using $10^6$ Monte Carlo samples drawn from a normal distribution with mean $8.89~\mu m$ and standard deviation $0.40~\mu m$, corresponding to the standard error of the mean thickness measured from four foil pieces. The 68$\%$ uncertainty interval gives $7\times10^{-6}<\Delta_{n3}^{th}<17\times 10^{-6}.$ $Z_{n3}$ is calculated to lie in the range $36 \Omega <|Z_{n3}|<95 \Omega$. This small value is intuitively consistent with the parallel-circuit picture: only a small current flows across the foil, so the resulting coupling is weak when the foil thickness spans multiple skin depths. The experimentally fitted values $\Delta_{13}=(10.8\pm0.3)\times10^{-6}$ and $\Delta_{23}=(8.6\pm0.2)\times10^{-6}$ both lie within the predicted $68\%$ interval. Given the strong sensitivity of the coupling to foil thickness and other geometric parameters, agreement at this level is consistent with and supports the model.

The remaining variation between the predicted and fitted coupling values likely arises from uncertainties in highly sensitive parameters, such as the foil thickness, where a difference of a few micrometres can result in an order-of-magnitude change in the coupling. The foil thickness is considered to be uniform for the purposes of the calculation. A more precise measurement of the foil thickness would reduce this uncertainty. Additionally, the material parameters such as conductivity may vary due to oxidation and other cavity differences. Small variations in the resonator mode structures will also lead to variations in the calculated mutual impedances.

\section{Conclusion}
We develop an equivalent circuit model to describe the coupling between three microwave cavity resonators interconnected by thin metallic foils. By introducing mutual coupling terms, the model accurately reproduces the measured interference features and antiresonant response. The derived coupling values agree with those extracted from the fitted transfer function, supporting the foil-mediated inter-cavity coupling model.

The demonstrated framework provides a basis for the design and analysis of microwave resonator networks requiring controlled weak coupling, antiresonant transmission, and enhanced phase response. Unlike localised probe- or aperture-mediated coupling, the foil-mediated interaction considered here is distributed across the shared conducting boundary between cavity fields, making it relevant to applications where the spatial structure of the coupled fields is important. This weak-coupling regime is particularly suited to controlled cancellation, rather than maximum power transfer. Future work may explore systematic parametric studies of foil thickness, alternate foil geometries, modal configurations, cryogenic implementations, and engineered constructive-coupling regimes. Collectively, these results establish metallic-foil-mediated mutual impedance as a practical and physically grounded coupling mechanism in multi-resonator systems.

\section*{Acknowledgements}
This work is supported by the Australian Research Council grants CE200100008, and DE250100933. This work is also supported by the Defence Science Centre, an initiative of the State Government of Western Australia.

We thank Eugene N. Ivanov for valuable assistance in experimental preparation.

\bibliographystyle{IEEEtran}
\bibliography{3cavitymodel.bib}

\appendices
\section{Transition Boundary Condition Derivation}
\label{derivation}

Across a thin material layer, the electric fields tangential to the surface on side 1 and side 2, $\tilde{E}_{t1}$ and $\tilde{E}_{t2}$, respectively, are given by \begin{eqnarray}
	\tilde{E}_{t1} &=& Z_{s} \tilde{J}_{s1} + Z_{t} \tilde{J}_{s2}, \quad \label{EZJ1} \\
	\tilde{E}_{t2} &=& Z_{t} \tilde{J}_{s1} + Z_{s} \tilde{J}_{s2}. \quad \label{EZJ2}
\end{eqnarray} $Z_{s}$ and $Z_{t}$ are the self and transfer impedances of the material layer, and $\tilde{J}_{s1}$ and $\tilde{J}_{s2}$ the surface current densities on the two sides. This comes from the fact that the boundary condition due to a large conductivity allows us to generally write, with little error \cite{collin2007foundations}, $\tilde{E}_{t}=Z_{surface} \tilde{J}_{s}$, where $Z_{surface}$ is the surface impedance. Solving for the surface currents in Eqs.~\eqref{EZJ1} and ~\eqref{EZJ2}, we arrive at\cite{COMSOL_Optics_UserGuide_5.5} \begin{eqnarray}
	\tilde{J}_{s1}=\frac{Z_{s}\tilde{E}_{t1}-Z_{t}\tilde{E}_{t2}}{Z_{s}^{2}-Z_{t}^{2}}, \quad \\
	\tilde{J}_{s2}=\frac{Z_{s}\tilde{E}_{t2}-Z_{t}\tilde{E}_{t1}}{Z_{s}^{2}-Z_{t}^{2}}. \quad
\end{eqnarray} Following the standard transition matrices one has \begin{equation}
\begin{bmatrix}
	\tilde{V}_{1}\\
	\tilde{I}_{1}
\end{bmatrix}
=
\begin{bmatrix}
	\cosh(\gamma d) & Z_0 \sinh(\gamma d) \\
	\frac{1}{Z_0} \sinh(\gamma d) & \cosh(\gamma d)
\end{bmatrix} \begin{bmatrix}
\tilde{V}_{2}\\
\tilde{I}_{2}
\end{bmatrix}
\end{equation} and \begin{equation}
\begin{bmatrix}
	\tilde{V}_{1}\\
	\tilde{V}_{2}
	\end{bmatrix}=
\begin{bmatrix}
	Z_{11} & Z_{12} \\
	Z_{21} & Z_{22}
\end{bmatrix} \begin{bmatrix}
\tilde{I}_{1}\\
\tilde{I}_{2}
\end{bmatrix},
\end{equation} for voltages and currents $\tilde{V}_{n}$ and $\tilde{I}_{n}$, respectively. Here, $d$ is the propagation foil thickness, and $\gamma=i \omega \sqrt{(\epsilon+\frac{\sigma}{\omega i})\mu}$ is the propagation constant (wave number) determined by the material permittivity, conductivity, and permeability. One can then solve for $Z_{s}$ and $Z_{t}$ by defining $Z_{s}=Z_{11}$ and $Z_{t}=-Z_{12}$ arriving at \begin{eqnarray}
Z_{s}=-i Z_{0} \frac{1}{tan(k d)}, \quad \\
Z_{t}=-i Z_{0} \frac{1}{sin(k d)}, \quad
\end{eqnarray} setting $\gamma = i k$. Finally, the characteristic impedance is given by \cite{collin2007foundations} \begin{equation}
Z_{0}=\sqrt{\frac{\mu}{(\epsilon+\frac{\sigma}{i \omega})}}=\frac{\omega \mu}{k}.
\end{equation} And thus we obtain the transition boundary condition \cite{COMSOL_Optics_UserGuide_5.5}.

\section{Complete Foil-Current Coupling Derivation}
\label{Appendix:correction}
\begin{figure*}[t]
    \centering
    \begin{minipage}{0.39\textwidth}
        \centering
        \includegraphics[height=2.6cm]{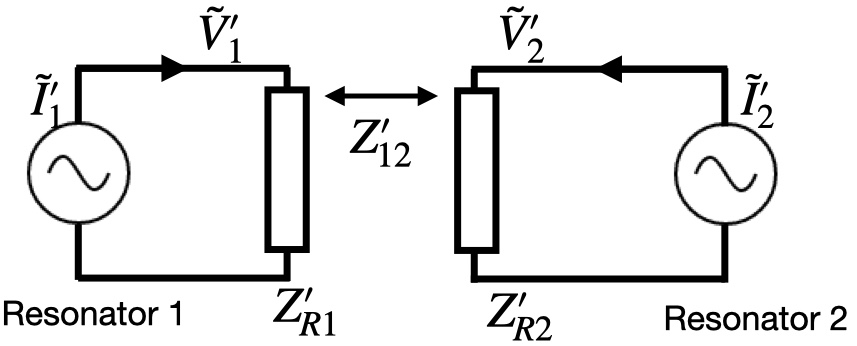}
        \par\smallskip{\normalsize (a)\par}
    \end{minipage}\hfill
    \begin{minipage}{0.59\textwidth}
        \centering
        \includegraphics[height=2.6cm]{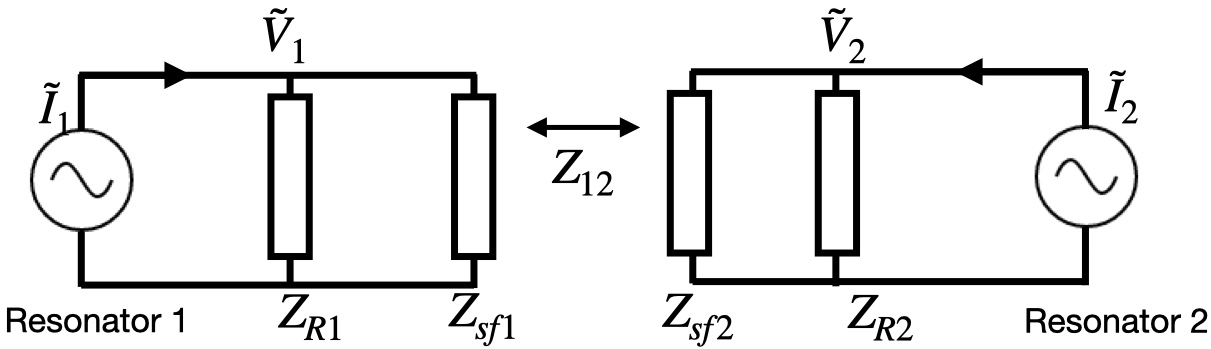}
        \par\smallskip{\normalsize (b)\par}
    \end{minipage}  
    \caption{ Equivalent circuits for a generic foil-coupled resonator pair. a) Compact lumped model with effective impedances $Z_{R1}'$, $Z_{R2}'$, and $Z_{12}'$. b) Separate-branch model where $Z_{Rj}$ excludes the foil contribution, $Z_{sfj}$ is the effective self-impedance of foil face j, and $Z_{12}$ is the mutual impedance between the two foil-current branches.}
    \label{fig:lumpedvsseperate}
\end{figure*}

In this appendix, indices 1 and 2 denote a generic coupled cavity pair and correspond to indices n and 3 in the main text. Thorough treatment of the circuit currents involved in the mutual coupling of two resonators via a shared surface such as a thin metallic foil requires splitting the current branches for that travelling along the foil and that which belongs to the rest of the resonator and does not directly participate. The approximate representation used in this paper for the simplicity of the transfer function is represented by Fig.~\ref{fig:lumpedvsseperate}(a). In this Appendix, the separate branch current representation shown in Fig.~\ref{fig:lumpedvsseperate}(b) will be treated carefully and shown to approach the lumped model in the small coupling, near-resonance limit.

\subsection{Self Impedance of the Foil}
First, we consider the self impedance of the foil $Z_{sf}$, where this is now separated from the impedance due to the other surfaces of the resonator. By considering only one resonator, or equivalently with no mutual coupling, the node equations are \begin{equation}
    \tilde{V}=Z_{R} \tilde{I}_{R} = Z_{sf} \tilde{I}_{sf},
\end{equation} where $Z_{R}$ is the parallel shunt impedance of the resonator excluding the contributions of the foil, with the corresponding current $\tilde{I}_{R}$. Similarly $\tilde{I}_{sf}$ is the current passing through the shunt foil self impedance. The relevant power equations are \begin{equation}
    P_{sf}=\frac{1}{2}\frac{|\tilde{V}|^2}{Z_{sf}^*},
    \label{sfp1}
\end{equation} with node voltage $\tilde{V}$, and \begin{equation}
    P_{sf}=\frac{1}{2}\int_{S_f}Z_s|\hat{n}\times \tilde{H}|^2 dS,
    \label{sfp2}
\end{equation} where $Z_s$ is the surface impedance of the foil, $\hat{n}$ is the unit vector normal to the foil surface, and $\tilde{H}$ is the magnetic field at the surface. The integral is over the surface of the foil $Sf$. Equating Eq.~\eqref{sfp1} and \eqref{sfp2}, and substituting $|\tilde{V}|=\sqrt{\frac{2U}{C}}$ with stored energy $U$, and capacitance $C$, we arrive at the expression for the foil self-impedance: \begin{equation}
    Z_{sf}=\frac{2U}{C \int_{S_f}{Z_s^*|\hat{n}\times \tilde{H}^*|^2dS}}.
\end{equation}

\subsection{Mutual Impedance of the Foil}
We now consider the interaction of two resonators to derive the mutual impedance. Let the two node voltages in Fig.~\ref{fig:lumpedvsseperate}(b) be \begin{equation}
    \tilde{V}_1 = \tilde{I}_{R1} Z_{R1} = \tilde{I}_{sf1}Z_{sf1}+\tilde{I}_{sf2}Z_{12},\\
    \label{seperatseqns2}
\end{equation}
\begin{equation}
    \tilde{V}_{2} = \tilde{I}_{R2} Z_{R2} = \tilde{I}_{sf2} Z_{sf2} +\tilde{I}_{sf1} Z_{12},
    \label{seperateeqns1}
\end{equation} where $\tilde{I}_{R1}$ and $\tilde{I}_{R2}$ are the currents travelling across the resonator 1 and 2, respectively, and $\tilde{I}_{sf1}$ and $\tilde{I}_{sf2}$ are the currents travelling across the foil on the sides of resonator 1 and 2, respectively. Using the same surface-current densities as in Eqs.~\eqref{En} and \eqref{E3}, we arrive at the same mutual power term: \begin{equation}
    P_{12}=Z_t\int_{S_f} Re(\tilde{J}_{s1}.\tilde{J}_{s2}^*)dS,
    \label{apppower1}
\end{equation} with the same definitions from Section \ref{couplingderivation}. Similar to Eq.~\eqref{diss2}, the mutual component of complex power on the foil can also be written as \begin{equation}
    P_{12}=Z_{12}Re(\tilde{I}_{sf1} \tilde{I}_{sf2}^*).
    \label{apppower2}
\end{equation} Eqs.~\eqref{apppower1} and \eqref{apppower2} yield \begin{equation}
    Z_{12}=\frac{Z_t \int_{S_f} Re(\tilde{J}_{s1}.\tilde{J}_{s2}^*)dS}{Re(\tilde{I}_{sf1}\tilde{I}_{sf2}^*)},
\end{equation} in terms of the current phasors.

By solving the matrix equation \begin{equation}
    \begin{pmatrix}
        \tilde{V}_1\\
        \tilde{V}_2
    \end{pmatrix}=\begin{pmatrix}
        Z_{sf1} & Z_{12}\\
        Z_{12} & Z_{sf2}
    \end{pmatrix} \begin{pmatrix}
        \tilde{I}_{sf1}\\
        \tilde{I}_{sf2}
    \end{pmatrix},
\end{equation} we obtain \begin{eqnarray}
    \tilde{I}_{sf1}&=&\frac{Z_{sf2} \tilde{V}_1}{Z_{sf1} Z_{sf2}-Z_{12}^2}-\frac{Z_{12} \tilde{V}_2}{Z_{sf1} Z_{sf2} - Z_{12}^2},\\
    \tilde{I}_{sf2}&=&\frac{Z_{sf1} \tilde{V}_2}{Z_{sf2} Z_{sf1}-Z_{12}^2}-\frac{Z_{12} \tilde{V}_1}{Z_{sf2} Z_{sf1} - Z_{12}^2}.
\end{eqnarray} For very small $|Z_{12}|<<|Z_{sf1}|,|Z_{sf2}|$, such as the presently studied case, we have \begin{eqnarray}
    \tilde{I}_{sf1} &\approx&\frac{\tilde{V}_1}{Z_{sf1}},\\
    \tilde{I}_{sf2} &\approx&\frac{\tilde{V}_2}{Z_{sf2}}.
\end{eqnarray}

Representing the surface- and equivalent foil-current phasors as $\tilde{J}_{sn}(r)=\vec{J}_{sn}(r) e^{i \psi_n}$ and $\tilde{I}_{sfn}=I_{sfn}e^{i\psi_n}$, where $\vec{J}_{sn}(r)$ is the phase-stripped modal current distribution which may be chosen real for the TM$_{010}$ mode and $I_{sfn}=|\tilde{I}_{sfn}|.$ $\psi_n$ is the foil-current phase, where the surface-current distribution and its corresponding foil-branch current carry the same phase. It follows that $\int_{S_f} Re(\tilde{J}_{s1}.\tilde{J}_{s2}^*)dS=cos(\psi_1-\psi_2)\int_{S_f} \vec{J}_{s1}.\vec{J}_{s2}dS$, while $Re(\tilde{I}_{sf1}\tilde{I}_{sf2}^*)=I_{sf1}I_{sf2} \cos(\psi_1-\psi_2).$ Therefore, the common phase factors cancel and we are left with \begin{equation}
    Z_{12}=Z_t\frac{\int_{S_f} \vec{J}_{s1}.\vec{J}_{s2}dS}{I_{sf1} I_{sf2}}.
\end{equation}
Defining the real modal-overlap factor \begin{equation}
    \mathcal{M}_{12}=\int_{S_f} Re((\vec{n}_1\times \vec{H}_1).(\vec{n}_2\times\vec{H}_2)^*)dS,
\end{equation} where $\vec{H_1}$, $\vec{H_2}$, are the phase-factored tangential magnetic fields at the respective foil side, and $\hat{n}_1$, $\hat{n}_2$, are the unit normal vectors at the respective foil side.
Evaluating the modal quantities near resonance, where $I_{sfn}=\frac{|\tilde{V}_n|}{|Z_{sfn}|}=\frac{1}{|Z_{sfn}|} \sqrt{\frac{2 U_n}{C_n}}$ we acquire the equivalent of Eq.~\eqref{unsubbedimpednace}: \begin{equation}
    Z_{12}=\frac{Z_t |Z_{sf1}| |Z_{sf2}|}{2} \sqrt{\frac{C_1 C_2}{U_1 U_2}}\mathcal{M}_{12}.
    \label{complexthorough}
\end{equation} The dissipative coupling involves taking the real part of \eqref{complexthorough}: \begin{equation}
    R_{12}=Re(Z_{12})=\frac{Re(Z_t) |Z_{sf1}| |Z_{sf2}|}{2} \sqrt{\frac{C_1 C_2}{U_1 U_2}} \mathcal{M}_{12}.
    \label{mutualresistancethorough}
\end{equation}

\subsection{Mapping Between the Separate-Branch and Lumped Models}
The two different models are shown to be similar in the small coupling limit. Primes denote the effective lumped quantities obtained after eliminating the explicit foil-current branches. Consequently, the primed impedances incorporate the foil-surface-integral contributions associated with the eliminated branches.

Solving Eqs.~\eqref{seperateeqns1} and \eqref{seperatseqns2} in addition to current conservation yields node voltages \begin{eqnarray}
    \tilde{V}_1=\frac{Z_{R1}(-\tilde{I}_1 Z_{12}^2 +\tilde{I}_2 Z_{12} Z_{R2}+\tilde{I}_1 Z_{sf1}(Z_{sf2}+Z_{R2}))}{-Z_{12}^2+(Z_{sf1}+Z_{R1})(Z_{sf2}+Z_{R2})},\\
    \tilde{V}_2=\frac{Z_{R2}(-\tilde{I}_2 Z_{12}^2 +\tilde{I}_1 Z_{12} Z_{R1}+\tilde{I}_2 Z_{sf2}(Z_{sf1}+Z_{R1}))}{-Z_{12}^2+(Z_{sf2}+Z_{R2})(Z_{sf1}+Z_{R1})}.
\end{eqnarray} In the small coupling limit $|Z_{12}|<<|Z_{sf1}|,|Z_{sf2}|$ this becomes \begin{eqnarray}
    \tilde{V}_1&=&\frac{Z_{R1}(\tilde{I}_2 Z_{12} Z_{R2}+\tilde{I}_1 Z_{sf1}(Z_{sf2}+Z_{R2}))}{(Z_{sf1}+Z_{R1})(Z_{sf2}+Z_{R2})},\\
    \tilde{V}_2&=&\frac{Z_{R2}(\tilde{I}_1 Z_{12} Z_{R1}+\tilde{I}_2 Z_{sf2}(Z_{sf1}+Z_{R1}))}{(Z_{sf2}+Z_{R2})(Z_{sf1}+Z_{R1})}.
\end{eqnarray}

The node equations for the lumped case are much simpler: \begin{eqnarray}
    \tilde{V}_1'&=&\tilde{I}_1' Z_{R1}'+\tilde{I}_{2}'Z_{12}',\\
    \tilde{V}_2'&=&\tilde{I}_2' Z_{R2}'+\tilde{I}_{1}'Z_{12}'.
\end{eqnarray} One can see that the voltage expressions for the lumped and separate equations take the same form via the substitutions 
\begin{equation}
    Z_{R}' = \frac{Z_{R}Z_{sf}}{Z_{sf} + Z_{R}},
    \label{lumpedz}
\end{equation} and \begin{equation}
    Z_{12}'=\frac{Z_{12}Z_{R1}Z_{R2}}{(Z_{sf1}+Z_{R1})(Z_{sf2}+Z_{R2})}=\frac{Z_{12}Z_{R1}'Z_{R2}'}{Z_{sf1} Z_{sf2}}.
    \label{lumpedapproximation}
\end{equation}
\begin{figure}
    \centering
    \includegraphics[width=1\linewidth]{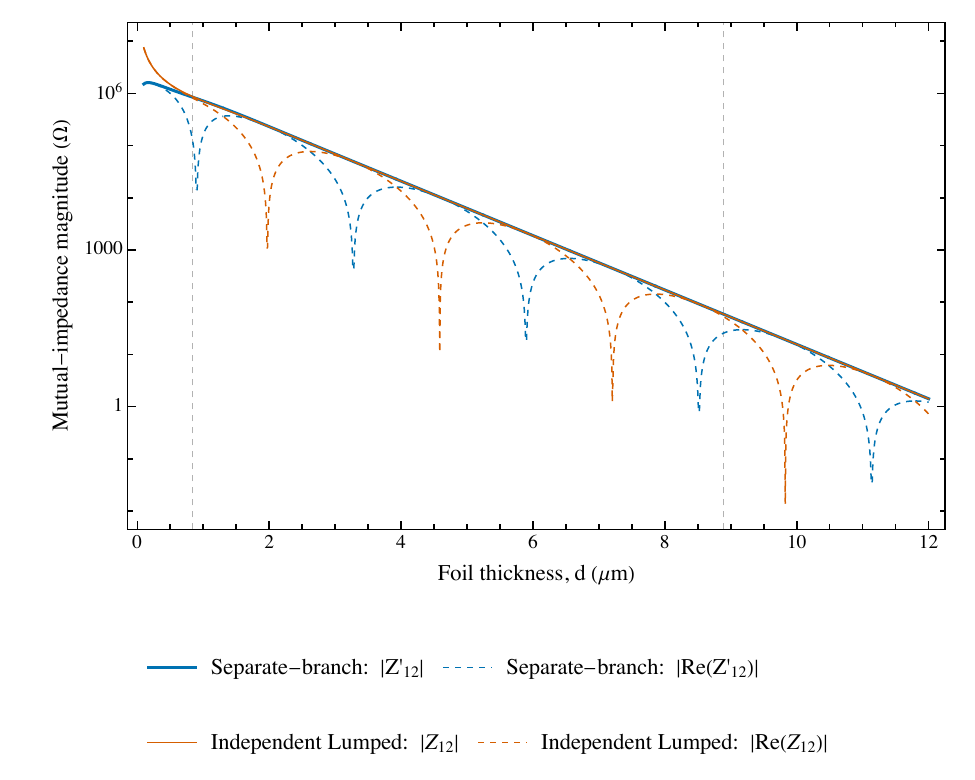}
    \caption{Comparison of the equivalent lumped mutual impedance obtained from the separate-branch formulation using Eqs.~\eqref{complexthorough} and \eqref{lumpedapproximation} (blue) and from the independent lumped formulation of Eq.~\eqref{mutualadmittance} (orange), as a function of copper foil thickness. Solid curves show the total coupling strength $|Z_{12}'|,$ which determines the normalised coupling coefficient $\Delta_{12},$ while the dashed curves show the magnitude of the dissipative component $|Re(Z_{12}')|.$ Agreement between the solid curves above approximately one skin depth validates the equivalent lumped mapping in the weak coupling regime. Differences between the dashed curves arise from the phase rotation introduced by the complex foil self-impedances and do not represent disagreement in mutual-impedance magnitude. The vertical dashed lines indicate copper skin depth at the operating frequency, $\delta=0.83$ $\mu$m, and the measured mean foil thickness, $d=8.9$ $\mu m$.}
    \label{fig:comparison}
\end{figure}
Equations \eqref{complexthorough}, \eqref{lumpedz} and \eqref{lumpedapproximation} establish the mapping between the physically separated foil-current circuit and the compact lumped circuit used in the main text. Under the relabelling $1 \xrightarrow{} n $ and $2 \xrightarrow{} 3$, $Z_{12}'\xrightarrow{}Z_{n3}'$, and the phase normalised effective quantity used in Fig.~\ref{circuitmodel} and Eq.~\eqref{mutual_coupling} is $Z_{n3}^{eff}=|Z_{n3}'|.$ Because the correction in Eq.~\eqref{lumpedapproximation} is complex, it must be applied to the mutual impedance before either the modulus or real part is evaluated. As shown in Fig.~\ref{fig:comparison}, the mapped separate branch and independently-derived lumped expressions agree in total coupling magnitude $|Z_{12}'|,$ above approximately one skin depth and begin to deviate below this scale.
Although both the magnitude and phase of the measured transmission are included in the fit, each coupling coefficient $\Delta_{n3}$ is represented as a real coupling strength parameter. In the present weak coupling regime, the phase of the corresponding complex mutual impedance is not independently resolved from the transmission phase reference and relative excitation phase. The fitted $\Delta_{n3}$ is therefore compared with $\frac{|Z_{n3}'|}{\sqrt{R_n' R_3'}}$ with $R_j'=Re(Z_{Rj}')$ the corresponding effective shunt resistance at resonance.

\section{Biography Section}
 
\vspace{11pt}

\begin{IEEEbiography}[{\includegraphics[width=1in,height=1.25in,clip,keepaspectratio]{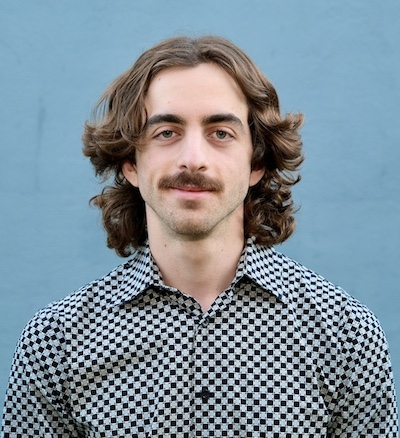}}]
{Michael T. Hatzon} (Student, IEEE) received the Bachelor of Philosophy (Honours) degree in Physics and Chemistry from the University of Western Australia (UWA), Perth, WA, Australia, in 2024. He is currently pursuing the Ph.D. degree in experimental physics with the Quantum Technologies and Dark Matter Research Laboratory at UWA.
His research focuses on precision microwave resonator engineering and advanced noise suppression techniques in measurement systems for tests of fundamental physics. His work includes both theoretical and experimental investigations of the scalar electric Aharonov-Bohm effect and the development of resonator-based detection architectures and analysis methods for axion and dark matter detection. In 2025, he received the International Union of Radio Science (URSI) Young Scientist Award for his contributions to precision microwave experiments and fundamental electromagnetic tests.
\end{IEEEbiography}

\vspace{11pt}

\begin{IEEEbiography}[{\includegraphics[width=1in,height=1.25in,clip,keepaspectratio]{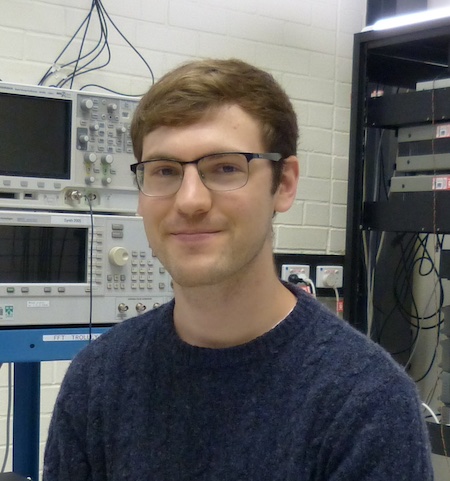}}]
{Graeme R. Flower} received the PhD degree from the University of Western Australia in 2022 working on improving axion dark matter detection experiments with magnetic materials. Currently he is a Research Associate working with the Quantum and Dark Matter lab at the University of Western Australia. His research interests include axion dark matter detection, quantum sensing, and precision metrology.
\end{IEEEbiography}

\vspace{11pt}

\begin{IEEEbiography}[{\includegraphics[width=1in,height=1.25in,clip,keepaspectratio]{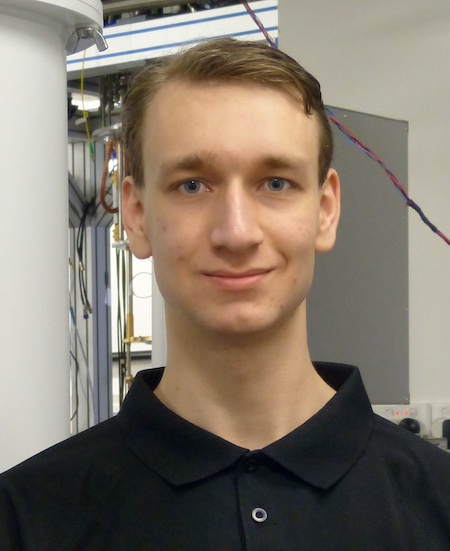}}]
{Robert C. Crew} received the Bachelor of Philosophy (Honours) degree in Electrical Engineering and Physics from the University of Western Australia (UWA), Perth, WA, Australia, in 2023. He is currently pursuing a Ph.D. degree in physics with the Quantum Technologies and Dark Matter research lab, at UWA. His research interests include frequency metrology, precision measurement, and axion dark matter detection.
\end{IEEEbiography}

\vspace{11pt}
\begin{IEEEbiography}[{\includegraphics[width=1in,height=1.25in,clip,keepaspectratio]{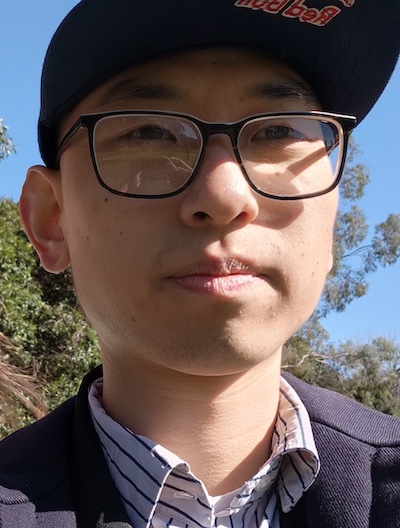}}]
{Ethan Han} received the Bachelor of science in theoretical physics from the university of Adelaide in 2025. He is currently Master by research student at Swinburne University of Technology, where his research focuses on using superconducting thin films to improve the quality factor of axion dark matter haloscopes. His work includes the development and characteriation of superconducting thin films for high frequency and high magnetic field applications.
\end{IEEEbiography}

\vspace{11pt}
\begin{IEEEbiography}[{\includegraphics[width=1in,height=1.25in,clip,keepaspectratio]{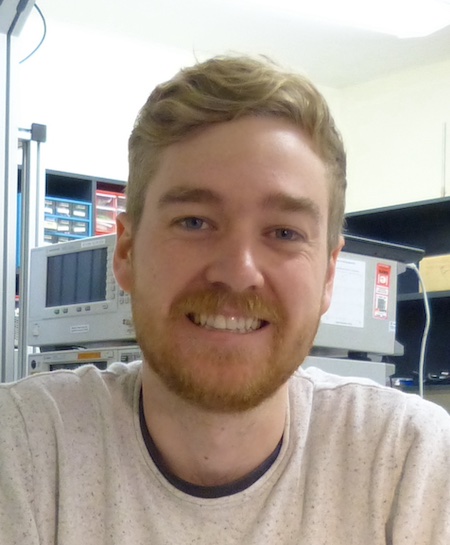}}]
{Ben McAllister} received the Ph.D. degree in physics from The University of Western Australia in 2019. He is currently an Australian Research Council DECRA Fellow at Swinburne University of Technology. His research interests include experimental searches for dark matter and other new phenomena, axion haloscopes, microwave resonators, superconducting devices, quantum sensing technologies, and deep underground physics.
\end{IEEEbiography}

\vspace{11pt}

\begin{IEEEbiography}[{\includegraphics[width=1in,height=1.25in,clip,keepaspectratio]{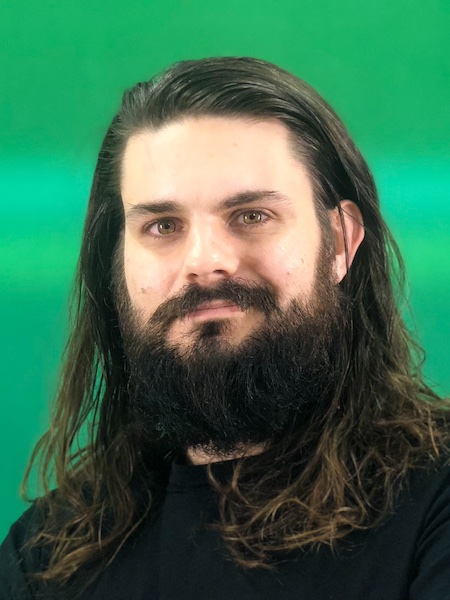}}]
 {Jeremy F. Bourhill} was born in Perth, Australia, in 1989. He received the B. Sc. degree (with honors) in physics from the University of Western Australia, Perth, Australia, in 2011, and the Ph.D. degree in physics from the University of Western Australia, Perth, Australia, in 2016.
From 2016 to 2019 he was a Research Associate in the Department of Physics, University of Western Australia, whilst from 2019 to 2021 he was a Research Engineer at IMT-Atlantique, in Brest, France. Since 2021 he has fulfilled the role of Research Fellow and now Lecturer at the University of Western Australia. His research interests include cavity magnonics, electromagnetic helicity, low-noise oscillators, quantum measurement, cryogenic systems, electromagnetics, microwaves, 3D printed superconductors and optomechanics.
\end{IEEEbiography}

\vspace{11pt}

\begin{IEEEbiography}[{\includegraphics[width=1in,height=1.25in,clip,keepaspectratio]{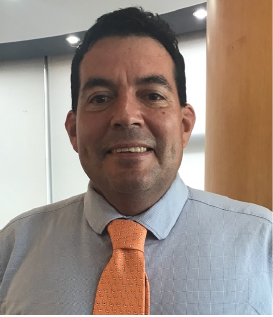}}]
{Michael E. Tobar} (Fellow, IEEE) received the Ph.D. degree in physics from The University of Western Australia (UWA), Perth, WA, Australia, in 1994.
He is currently a Professor of Physics with the Department of Physics at the University of Western Australia and leads the Quantum Technologies and Dark Matter Research Lab at UWA. His research group is part of the ARC Centre of Excellence for Dark Matter Particle Physics His broad research interests encompass the disciplines of frequency metrology, precision and quantum measurements, low temperature, condensed matter, and quantum physics. Over his career he has developed a variety of precision and quantum measurement tools in his laboratory allowing the investigation of many areas of Physics and Engineering. In particular he is interested in developing new precision technologies to undertaking precise tests of fundamental physics and has also adapted such technology to the commercial sector, which includes 13 patents in the area of precision radar and detectors. In the area of fundamental physics his work includes low energy precision searches for signs of quantum gravity, dark matter, Lorentz invariance violations and searches for high frequency gravitational waves. He also leads the ORGAN axion Dark Matter detector collaboration co-funded by both Centres. Also, in 2019 his group become an official collaborator of the Axion Dark Matter eXperiment (ADMX) situated at the University of Washington. Prof. Tobar has co-authored more than 350 refereed journal publications and has won numerous national and international awards for his work. Furthermore, he is a Fellow of the IEEE, the Australian Academy of Science, the Australian Academy of Science Technology and Engineering, and the Australian Institute of Physics. He was elected as a Voting Administration Committee (Adcom) Member for IEEE Ultrasonics, Ferroelectrics and Frequency Control Society for a three-year term from 2016 to 2018. From 2009 to 2014, he received the Laureate Fellowship from the Australian Research Council, the top nation-wide research fellowship. Other significant awards include the 2006 Boas Medal, the 2012 Alan Walsh Medal and the 2024 Harrie Massey medal from the Australian Institute of Physics, the 2009 Barry Inglis Medal from the National Measurement Institute, the 2010 Western Australian Scientist of the year from the Department of Commerce, WA, the 2014 W.G. Cady Award and the 2025 The C.B. Sawyer award from IEEE, the 2025 European Frequency and Time Award, and the 2014 Clunies-Ross Award from the Australian Academy of Science and Technology. He also received a citation from the Australian Learning and Teaching Council for inspiring research students to reach their full potential and transform to successful research scientists through participation in ground-breaking research.

\end{IEEEbiography}

\vfill

\end{document}